# The polarizing impact of numeracy, economic literacy, and science literacy on attitudes toward immigration

Lucia Savadori, Giuseppe Espa, and Maria Michela Dickson

*Department of Economics, University of Trento (Italy)*

**Abstract**

Political orientation polarizes the attitudes of more educated individuals on controversial issues. A highly controversial issue in Europe is immigration. We found the same polarizing pattern for opinion toward immigration in a representative sample of citizens of a southern European middle-size city. Citizens with higher numeracy, scientific and economic literacy presented a more polarized view of immigration, depending on their worldview orientation. Highly knowledgeable individuals endorsing an egalitarian-communitarian worldview were more in favor of immigration, whereas highly knowledgeable individuals with a hierarchical-individualist worldview were less in favor of immigration. Those low in numerical, economic, and scientific literacy did not show a polarized attitude. Results highlight the central role of socio-political orientation over information theories in shaping attitudes toward immigration.

# The Polarizing Impact of Numeracy, Economic Literacy, and Science Literacy on Attitudes Toward Immigration

**Introduction**

People's political orientation has been shown to polarize public opinion of more educated individuals on controversial issues such as climate change or stem cell research. This tendency has been termed "funnel pattern" (Drummond & Fischhoff, 2017) since the gap between the beliefs of political conservatives and political liberals widens as the level of education increases, mimicking a funnel shape. The more educated and scientifically literate individuals sustain the need to contrast climatic change more if they are left-oriented than if they are right-oriented. In contrast, the less educated and lower scientifically literate individuals do not show such a diverging trend in their opinions (Brewer, 2012; Drummond & Fischhoff, 2017; Hamilton, 2008, 2011; Hamilton et al., 2014; Hamilton, Hartter, Lemcke-Stampone, et al., 2015; Hamilton & Lemcke-Stampone, 2014; Kahan et al., 2012; Malka et al., 2009; McCright et al., 2016; McCright & Dunlap, 2011).

The political polarization of opinions indicates that most educated citizens are not those with more moderate opinions; instead, they are those with more extreme opinions. However, the most extreme opinions are not based on knowledge but are guided by their political orientation. Political polarization has important implications for the impact of information policies designed to increase citizens' capability to make informed choices on socially relevant issues. Paradoxically, implementing information and educational policies would increase people's polarization on extremist and opposing views producing a potential increase in social conflict.

Climate change and environmental issues are not the only controversial topics for which the funnel pattern has been observed. The same pattern was also detected on opinion about vaccines (Hamilton, Hartter, & Saito, 2015; Joslyn & Sylvester, 2017), support for embryonic stem cell research (Drummond & Fischhoff, 2017; Nisbet & Markowitz, 2014), and opinion toward the Big Bang and human evolution (Drummond & Fischhoff, 2017). There was, however, no interaction between education and political or religious identity on nanotechnology and genetically modified food. The authors advanced the explanation that these issues generated controversy, but they did not become part of the most significant social conflicts in America (Drummond & Fischhoff, 2017).

Most studies on the polarizing effect of education have focused mainly on environmental or scientific topics ignoring other issues. Instead, citizens' knowledge and education are also critical for sound decision-making on societal issues, such as immigration policies. Immigration is a very controversial issue in Europe (European Commission, 2019). Consequently, Europeans opinion on this topic strongly determines the European parties' political agenda (Dennison & Geddes, 2019; Goodman, 2019). Therefore, it is crucial to understand the moderating effect of political orientation on the relationship between education and citizens' opinion toward immigration.

Most of the data from previous studies showing the funnel pattern come from surveys of American citizens accustomed to a political arena characterized by strong bipartisanship (Republican vs. Democrats). The European political arena, instead, is complex and characterized by numerous parties, some of which do not identify as right- or left-wing (Hix et al., 2006; Manucci & Weber, 2017; Roccato et al., 2020; Schakel, 2018). Studies replicating the funnel pattern in the European political arena are missing but could be helpful in understanding social processes.

**Opinion Toward Immigration**

Immigration has been steadily rising in all European countries until the mid-1990s (OECD, 2020). More than a third of Europeans consider immigration the most important issue facing the European Union (European Commission, 2019). Therefore, it is of no surprise that immigration has also become an increasingly central political issue on the agenda of anti-immigration parties (Dennison & Geddes, 2019).

The debate on the benefits and the risks of immigration might deal with economic issues. On the one hand, immigration can be south to benefit the economy, providing companies with skilled workers, and relieving tension on the tax-funded pension system threatened by the lack of population growth. On the other hand, citizens might fear that immigrants may take jobs away from local workers and take more from the government in the form of social services than they give back in taxes. Some studies have concluded that the fears about the economic effects of competition on the labor market among low-skilled workers and blue-collar workers are at the root of anti-immigration feelings (Scheve & Slaughter, 2001).

Given that much of the debate around immigration is about economic issues, it is essential to explore how economic knowledge affects people's opinion toward immigration and how political orientation can moderate this relationship.

Economic knowledge, also called economic literacy, is the personal knowledge about basic economic concepts, such as markets and prices, supply and demand, money and inflation, economic institutions, labor markets, income, etc. (Walstad et al., 2013). Economic literacy must not be confused with financial literacy. Financial literacy is the personal knowledge about concepts of financial management, budget, and investment, such as risk diversification, interest compounding, mortgages, other debt instruments, etc.. In the present study, we have explored the effect that individuals' economic literacy has on their attitude towards immigration and how individuals' cultural worldview orientation can moderate this effect. We have been looking for empirical evidence that those with more economic knowledge are also more polarized by their worldviews in their opinion on immigration than those with less economic knowledge.

Among the factors predicting opposition against immigration, the most studied is, without doubt, education. Those with a higher education level have more positive attitudes toward immigrants and are more inclined to favor tolerant immigration policies (Hainmueller & Hiscox, 2007; Scheve & Slaughter, 2001). Education is strongly associated with numeracy (Peters et al., 2010; Zebian & Ansari, 2012) and science literacy (Drummond & Fischhoff, 2017; National Science Board, 2020). Numeracy refers to individuals' capacity to comprehend and use quantitative information (Peters et al., 2007). Science literacy is the knowledge of basic scientific facts (National Science Board, 2020). Since both science literacy and numeracy were used as predictive factors in previous studies that identified a polarizing pattern (Drummond & Fischhoff, 2017; Kahan et al., 2012), we also included them in the present study. However, our decision to include science literacy and numeracy also stems from other considerations. The arguments against immigrants could be associated with the immigration rate, which might be considered too high, or crime rates, which could be seen as associated with immigration rates. Mastering the quantitative and statistical aspects is, therefore, essential to have an informed opinion. Other arguments against immigrants might be related to the fact that they can be vehicles for viruses and diseases. Having scientific knowledge, in this case, would allow people to have a more well-founded opinion. Other arguments in favor of immigration might refer to its importance for science and technology, as citizens might believe that qualified workers (i.e., engineers and scientists) need to be imported from outside to sustain their country's technological development.

It has been noted that a large part of the link observed between education and attitudes toward immigrants is driven by differences between individuals regarding their cultural values and beliefs. For example, most educated respondents are significantly less racist and value cultural diversity more than their counterparts (Hainmueller & Hiscox, 2007). Seemingly, drawing from the European Social Survey data, it was found that conservation values were associated with more significant immigration threats than universalism values (Davidov et al., 2020). Likewise, politically conservative Europeans perceived a greater threat from immigration than other Europeans (Czymara, 2020)

What is unknown is how education shapes beliefs on immigration, depending on individuals' political orientation. This study explored whether the citizens' level of numeracy, science literacy, and economic literacy can shape their attitudes toward immigration and how cultural worldviews moderate this relationship, producing a polarizing pattern.

Most studies on the polarization of opinions used a question on political orientation to label people as right-wing or left-wing. Instead, we preferred to use a measure of cultural worldviews (Kahan et al., 2011). The reasons are two. First, the standard question to measure political orientation does not fit well into the European political scene. The political question asks people to define themselves based on two dimensions: republican vs. democrat and liberal vs. conservative (Hamilton, Hartter, & Saito, 2015; Smith et al., 2017). Although these two dimensions are adequate to map American citizens' political orientation, we believe that they are not adequate for European citizens. Indeed, the European political landscape is characterized by a great fragmentation of opinions into small parties that often have ideas transversal to the two dimensions (Hix et al., 2006; Manucci & Weber, 2017; Roccato et al., 2020; Schakel, 2018). For example, the Green Party gathers both right-wing and left-wing voters, both liberals and conservatives, united by the idea of preserving the environment. Populist parties, likewise, have within them both right-wing (e.g., anti-immigrant) and left-wing (e.g., guaranteed minimum income) ideologies, as well as both conservative (e.g., NO TAV movement) and liberal (e.g., drug liberalization) ideologies. Second, some authors (Van Bavel & Pereira, 2018) argue that political preference is only an external manifestation of a deeper set of values, and it has been claimed that human values should be considered the cause of the cause of opposition to immigration (Davidov et al., 2020).

Among human values, we investigated cultural worldviews. Indeed, previous literature has shown that cultural worldviews capture the political polarization of opinions of the most educated (Kahan et al., 2012). Moreover, worldviews explain the attitudes of people in the face of threats. To the extent that citizens view immigration as a potential societal threat, then worldviews should explain individual differences in attitudes toward immigration well.

According to cultural theory, people perceive risk when their cultural worldview is challenged (Douglas & Wildavsky, 1982). The theory includes four worldviews of societal organization: hierarchism, egalitarianism, individualism, and fatalism. In this study, following prior work (Kahan et al., 2011, 2012), we used a cultural cognition approach that defines two dimensions: hierarchy vs. egalitarianism and individualism vs. communitarianism. People with a hierarchical worldview prefer a hierarchical society clustered according to well-defined differences among groups identified by gender, race and class (e.g, "It seems like blacks, women, homosexuals and other groups don't want equal rights, they want special rights just for them"). On the opposite side, people with an egalitarian worldview prefer a society where minorities have equal rights and inequality is eradicated (e.g., We need to dramatically reduce inequalities between the rich and the poor, whites and people of color, and men and women"). People with an individualist worldview prefer a society where individuals will secure their well-being without assistance or interference from society (e.g., "The government interferes far too much in our everyday lives"). On the opposite pole, people who hold a communitarian worldview think that the government has the responsibility for collective welfare and the power to override individual interests (e.g., "Sometimes government needs to make laws that keep people from hurting themselves").

**The Present Research**

We examined how individual numeracy skills, science literacy, and economic literacy influence the opposition to immigration and how cultural worldviews moderate this relationship. It was expected that hierarchical-individualists would show a less favorable attitude toward immigration than egalitarian-communitarians and that this gap would increase as individual economic, scientific and numerical knowledge increased, replicating the funnel pattern already found for climate change.

There are several innovative aspects of this study. First, while previous studies have mainly examined American citizens' views, this study extends the scope to include European citizens, who have political backgrounds that are less bipartisan. Second, while previous studies have predominantly examined the views on scientific issues, this study examines a social issue, immigration, perhaps even more important for the potential repercussions on society of extreme radicalized positions. Third, while previous studies have predominantly used the level of education as a proxy for objective knowledge, in this study, knowledge is measured directly, thus more accurately, through the measurement of numeracy, science literacy, and economic literacy.

## Method

**Sample**

A representative sample of citizens of the city of Trento (Northern Italy) participated in the study. Data were collected from March 1 to April 30, 2019. The reference population of the survey was selected from the municipal register updated to January 1, 2019. The reference population included all adult citizens between 18 and 80 years of age of both sexes, resident in the city's territory, for a total of 90,051 units. Homeless and nomads were excluded from the reference population.

A stratified sampling method was adopted to select the sample: the population was divided into homogeneous, non-overlapping strata (groups) based on known stratification variables available for all units. The variables adopted for the stratification were the following: gender (female, male), date of birth (from 18 to 35 years, from 36 to 55 years, and from 56 to 80 years), and the district of residence (South-West, Centre-North, North-East) (See Figure1S in the Supplementary file). The population resulted in a multivariate stratification with a total of 18 strata (see Table 1S in the Supplementary file).

The allocation of the sample in each stratum was proportional to the stratum size, with a minimum number of units in each stratification cell $n_h = 5$ and a maximum $n_h = N_h$. When $n_h < 5$, the allocation is forced to $n_h = 5$ and when $n_h = 5$, the stratum is censused. We used the **R** package **sampling** (Tillé & Matei, 2016). The selection criterion in each stratum has been the simple random sampling without replacement, with inclusion probability for the $h-th$ stratum equal to $\pi_h = \frac{n_h}{N_h}$. The obtained sample was composed by 2,008 adult inhabitants, with a sampling fraction equal to 2%. This is a notable issue since, in national social surveys, the sampling fraction rarely exceeds 0.1% (see Table 2S of the Supplementary file for the allocation of the selected sample).

Participation was voluntary. In total, 551 people from the city of Trento completed the survey. The overall response rate was 27% (551 questionnaires completed out of 2,008 administered), a remarkable result compared to opinion surveys conducted in similar ways (see Table 3S in the Supplementary file for the sample response rate according to the stratification variables).

The sample of 551 respondents included 281 males (51%) and 270 females (49%) with an average age of 52.2 years (SD = 16.4). More than a third of the respondents (38.3%) had completed at least a Bachelor's degree, whereas 44.1% had completed at least a high-school diploma, and the remaining (17.6%) had completed lower levels of educations (or had no certificates).

Before proceeding with the analysis, we corrected the data for non-responses (Särndal & Lundström, 2005) to produce unbiased estimates. To this aim, a calibration estimator was implemented (Deville & Särndal, 1992), forcing the calibration on the three stratification variables totals. The implementation was carried out through the `R` package `survey` (Lumley, 2020).

**Materials and Procedure**

The survey was conducted through a questionnaire administered with CAWI/CATI methodologies through LimeSurvey's web application. Participants were contacted by letter and invited to participate autonomously in the online survey or contact the research group by phone to fill out the form.

The questionnaire was composed of several sections. At the beginning of each section, there was a short introduction. In the section dedicated to the measurement of opinion toward immigration, instructions read: "There are different opinions on immigrants living in Italy. By "immigrants" we mean people from other countries who come to settle in Italy. Here below we will ask you some questions about your opinions". In the section dedicated to measuring the cultural worldviews, instructions read: "In this section, we will ask you a series of questions about your socio-economic opinions. Please indicate your degree of agreement or disagreement for each statement". In the sections dedicated to the measurement of numeracy, economic literacy, and science literacy, it was repeated each time: "The questions in this section serve to measure your familiarity with [mathematics and probability; economic phenomena; scientific subjects]. The questions do not in any way measure intelligence but rather the habit you have formed over time to use your knowledge to perform mathematical and probabilistic calculations. You may not feel confident in answering some questions, but we invite you to find the answer that seems to be correct". At the end of the questionnaire, participants answered some demographic questions on gender, age, place of residence, and education.

**Cultural worldviews.** To measure cultural worldviews, we used the short version of the cultural worldviews scale proposed by Kahan et al. (2011). The scale consisted of 12 statements, and the participant was asked to report the degree of agreement with each of them. The statements have been designed to reflect two underlying bipolar dimensions (hierarchy vs. egalitarianism and individualism vs. communitarianism). For all items, participants indicated the degree of agreement or disagreement on a five-point scale (1 = completely agree; 5 = completely disagree). This scale was used successfully in European populations: both U.K. (Marris et al., 1998) and Dutch populations (Poortinga et al., 2002; Steg & Sievers, 2000). A reliable composite overall index ($\alpha$ = .73) expressing the degree of hierarchical-individualist worldview was obtained by averaging all the responses.

**Opinion on immigration.** Opinion on immigration was measured by asking participants to express their agreement with ten statements from the General Social Survey 1972-2014 (Smith et al., 2017) and answer three questions about risk perception (see Appendix). The ten statements from the General Social Survey asked about opinions on several aspects of immigration. Participants answered on a five-point scale (1 = completely agree; 5 = completely disagree). The risk perception questions asked about the risks and benefits of immigration for the Italian society as a whole (Finucane et al., 2000; Skagerlund et al., 2020) and the extent to which thinking about immigration was associated with positive or negative emotions as a measure of affective attitude, a component of risk perception (Markowitz &

Slovic, 2020; Peters & Slovic, 2007; Slovic et al., 2007). Participants answered on a five-point scale, both the risk and benefit perception questions (1 = not at all risky/beneficial; 5 = extremely risky/beneficial) and the emotion question (1 = very negative; 5 = very positive). We gave the option of answering: "I do not know / I do not answer." All items were reverse coded so that higher values corresponded to a more favorable attitude toward immigration. Responses were grouped in a reliable ($\alpha$ = .94) average index reflecting the degree of tolerance toward immigration.

**Numeracy.** Five questions from Weller et al. (2013) were used to measure objective numeracy (see Appendix). Each question had four possible answers, but only one was correct. To compute the numeracy index, we calculated the number of correct answers. Some of the items had to be adapted to the cultural context. We gave the option of answering: "I do not know / I do not answer."

**Economic literacy.** We used 12 questions from the Test of Economic Knowledge (Walstad et al., 2013) to measure economic literacy. The items n.8, 9, 13, 15, 17, 23, 25, 26, 30, 41, 42, 44 of the original test by Walstad et al. (2013) were used. Individual score on economic literacy was computed calculating the number of correct answers.

**Science literacy.** The questions to measure scientific knowledge were drawn from the Science & Engineering Indicators (National Science Board, 2016) (See Appendix). They are ten statements to which participants replied "true" or "false." The statements span from clinical aspects (e.g., antibiotics kill viruses, as well as bacteria) to biological aspects (e.g., it is the paternal gene that determines whether the child will be a male or a female) but also technological aspects (e.g., lasers work by focusing sound waves). The scientific knowledge index was calculated, considering the number of correct answers.

# Results

**Analytical Strategy**

To test whether cultural polarization was greater among respondents with greater knowledge, we fit a model predicting participants' opinion toward immigration (coded such that higher values represent a more positive attitude) as a function of measures of numeracy, economic literacy, science literacy, and a new compound measure of *total literacy* representing the mean of the aggregated knowledge measures. In each model, we estimated the direct effect of each explanatory variable on attitudes toward immigration as well as one interaction term—each interaction term combined worldview orientation with one of the four knowledge measures.

In the models, the response variable (opinion toward immigration) was computed for each individual $i$, as the mean of the scores of the ten statements from the General Social Survey and the three questions about risk perception. Hence, the estimated model in all four scenarios is $\hat{y}_i = b_0 + b_1 x_i + b_2 z_i + b_3 x_i \times z_i$, where $z$ is a dichotomous variable created to label individuals as egalitarian-communitarians or hierarchical-individualists. To create the $z$ variable, we averaged individual responses to the 12 statements of the cultural worldviews scale (say $w$) such that higher scores corresponded to a higher hierarchical-individualist orientation. The cut-off was fixed at the median level of 2.5, such that:

$$z_i = \begin{cases} 1 \text{ if } \bar{w}_i > 2.5 \text{ (hierarchical} - \text{individualist)} \\ 0 \text{ if } \bar{w}_i < 2.5 \text{ (egalitarian} - \text{communitarian)}. \end{cases}$$

Using this threshold, we had 214 cases for $z = 0$, and 334 cases for $z = 1$. The estimated prediction equation assumes the form $\hat{y}_i = b_0 + b_1 x_i$, if $z = 0$, and $\hat{y}_i = (b_0 + b_2) + (b_1 + b_3)x_i$, if $z = 1$.

The explanatory variable $x$ represents, in the order in which the models are presented, respectively numeracy, economic literacy, science literacy, and total literacy. Each $x$ is the sum of the correct answers given by each individual to the questions of each knowledge test. Using the sum is an ideal strategy given to the presence of "I don't know/I don't answer" as a possible response. Table 1 shows the results of the multiple regression models for the four considered explanatory variables.

---------------

Insert Table 1 here

-------------

**Main Effects**

Examining Table 1 it is worth noting that all the parameter estimates ($b_1$) of the $x$ variables have a positive sign denoting a positive association between the opinion toward immigration and each of the knowledge measures: numeracy, economic literacy, science literacy, and total literacy. The direct effects on the opinion toward immigration predicted by the model, by controlling for the other predictors, are stronger for numeracy and science literacy than economic literacy. In all cases, however, the associations are strongly significant (all *p-values* < .001). Participants' opinions toward immigration are significantly predicted by their score on the numeracy test, the science literacy test, and the economic literacy test. Participants scoring higher on these knowledge measures showed a more favorable attitude toward immigration.

The situation is different for the dummy variable $z$ (worldviews) for which the regression coefficients are not significant (see $b_3$-row). The issue does not deserve special attention because variable $z$ is contained in the interaction term that assumes, on the contrary, crucial importance, as discussed hereafter. It is sufficient to note that a non-significant regression coefficient for $z$ means that an individual classified as hierarchical-individualist, with no numerical, economic, scientific, and total literacy skills, has, on average, the same opinion toward immigration as another individual classified as egalitarian-communitarian. In other terms, the lack of these skills makes the contribution of the z variable (worldview) indifferent to the model predicting opinion towards immigration.

**Interactions With Worldviews**

The $b_3$-row in Table 1 reports the estimates of the regression coefficients for the interaction among the *dummy* z variable (worldviews) and each explanatory variable used in the four distinct models. All the coefficients have negative signs, and all interaction terms are significant. The reverse association is strongly significant (*p-value* < 1‰) for numeracy and total literacy. However, it is also significant (*p-value* < 5%) for economic literacy and science literacy. Therefore, we do not eliminate, even if not significant, the main effect of z in all four estimated models because, as said before, the z appears in the significant interactions. Moreover, the slope of the relationship among the response variable $y$ and each explanatory variable $x$ changes when we go from $z = 0$ to $z = 1$. It means that the two lines are not parallel, as shown in Figure 1.

---------------

Insert Figure 1 here

-------------

Individuals classified as egalitarian-communitarian increase their positive attitude towards immigration as their skill in numbers, economics, and science, increase. On the contrary, individuals classified as hierarchical-individualists do not increase their favorable attitude toward immigration as their competence on these topics increase, showing on average an opinion toward immigration basically flat. Roughly speaking, hierarchical-individualist individuals show a strong homogeneity in their opinions on immigration, while egalitarian-communitarian individuals demonstrate a relevant heterogeneity in their opinions on immigration, as explained by the four competence measures. As an example, and only for numeracy, we report in Table 2 the mean values of the response variable $y$ corresponding to the different numeracy levels for the egalitarian-communitarian individuals ($z = 0$) and for hierarchical-individualist individuals ($z = 1$).

--------------

Insert Table 2 about here

----------------

## Discussion

Previous studies found that political polarization on controversial issues, such as climate change, was greater among individuals with more education and scientific knowledge. Here, we examined the same funnel-shaped pattern, on attitudes towards immigration, in a representative sample of citizens of a southern European city. Overall, we confirmed that cultural polarization, measured by cultural worldviews, was greater among individuals with greater knowledge, measured by numerical, economic, and scientific literacy tests. We have observed that a hierarchical-individualistic worldview was significantly associated with less favorable attitudes toward immigration than an egalitarian-communitarian worldview, and this gap was greater for individuals with greater numerical skill, greater economic literacy, and greater scientific literacy.

These results are consistent with previous research that has found a political polarization of American beliefs on scientific issues. While previous studies have mainly used data from American samples, we have collected the opinions of a representative sample of citizens from a southern European city. Southern Europe (Italy, Greece, Spain) is at the forefront of unskilled migration flows from Africa, so it is a perfect field to study citizens' attitudes toward immigration. We also used a cultural worldview measure instead of a political orientation measure, demonstrating that cultural values can be an ideal substitute for measuring polarization effects in areas such as Europe, where there is no strong political bipartisanship. Besides, we have demonstrated that the funnel-shaped model is not limited to scientific issues but also extends to social issues, such as immigration. Finally, in addition to mathematics and scientific knowledge, already measured in previous studies, we have extended the concept of knowledge to economic phenomena, which we believe is very important to have an impartial evaluation of a phenomenon like immigration.

The funnel pattern highlights a paradoxically negative aspect of being more knowledgeable. It would show that more knowledge corresponds to greater radicalization of opinions and fears along value and cultural dimensions. This contradicts the idea that providing more information and education (statistical, economic, and scientific knowledge) to citizens is a way to reduce social conflict.

Why are more educated people also more politically polarized in their views? A motivated reasoning account is usually applied to explain the polarized views' pattern. Motivated reasoning would motivate people to selectively seek, elaborate, and recall information in a way that supports their a priori beliefs (Kunda, 1990). Since more educated people also tend to seek more information than less educated people, the funnel pattern would arise from the combination of increased information processing coupled with motivated reasoning (Drummond & Fischhoff, 2017; Kahan et al., 2012; Martin & Desmond, 2010). Biased motivated reasoning might reduce scientific and objective messages' effectiveness and intensify opposing positions' crystallization, thus increasing social tension (Corner et al., 2012). Motivated reasoning is of particular concern in the context of politicized science (Anglin, 2019), but it might be as equally critical for another controversial issue, such as immigration.

Our result that greater knowledge is associated with greater cultural polarization is consistent with the motivated reasoning explanation, but it was not within the study's scope to explore the causal mechanisms of the funnel pattern. Subsequent studies should investigate aspects such as group identity and understand its relationship to the cultural polarization of opinions and motivated numeracy and seek to explain attitudes' political polarization.

**Table 1:** *Linear regressions results predicting opinion toward immigration from x (numeracy, economic literacy, science literacy, and total literacy), z (egalitarian-communitarians vs. hierarchical-individualists), and the interaction between x and z.*

|  | Explanatory variable | | | |
| --- | --- | --- | --- | --- |
|  | Numeracy | Economic literacy | Science literacy | Total literacy |
| $b_0$ | 3.08494*** | 2.97800*** | 2.55780*** | 2.44528*** |
|  | (0.15734) | (0.17234) | (0.26509) | (0.24591) |
| $b_1$ | 0.14975*** | 0.06626*** | 0.11893*** | 0.05370*** |
|  | (0.04478) | (0.01804) | (0.03010) | (0.01135) |
| $b_2$ | -0.12563 | -0.21028 | 0.11541 | 0.27997 |
|  | (0.18495) | (0.20929) | (0.31213) | (0.28918) |
| $b_3$ | -0.15649** | -0.04721* | -0.08711* | -0.04327** |
|  | (0.05335) | (0.02209) | (0.03580) | (0.01348) |
| *RSE* | 0.5720 | 0.5693 | 0.5678 | 0.5643 |
| $R^2$ | 0.2539 | 0.2608 | 0.2647 | 0.2738 |
| *Adjusted* $R^2$ | 0.2492 | 0.2561 | 0.2600 | 0.2692 |
| *F* | 53.55*** | 55.51*** | 56.64*** | 59.32*** |

*Note*. SEs are shown in parentheses. RSE = Residual standard error.;
***p-value < 0.000, **p-value < 0.001, *p-value < 0.050.

**Table 2**: *Mean values of the variable opinion on immigration for the five levels taken by the explanatory variable numeracy, separately for egalitarian-communitarian individuals (z = 0) and for hierarchical-individualist individuals (z = 1).*

| $x$ = numeracy | $z = 0$ | $z = 1$ |
|---|---|---|
| 0 | 2.596 | 3.060 |
| 1 | 3.122 | 2.986 |
| 2 | 3.324 | 3.035 |
| 3 | 3.668 | 2.841 |
| 4 | 3.618 | 2.930 |
| 5 | 3.906 | 3.143 |

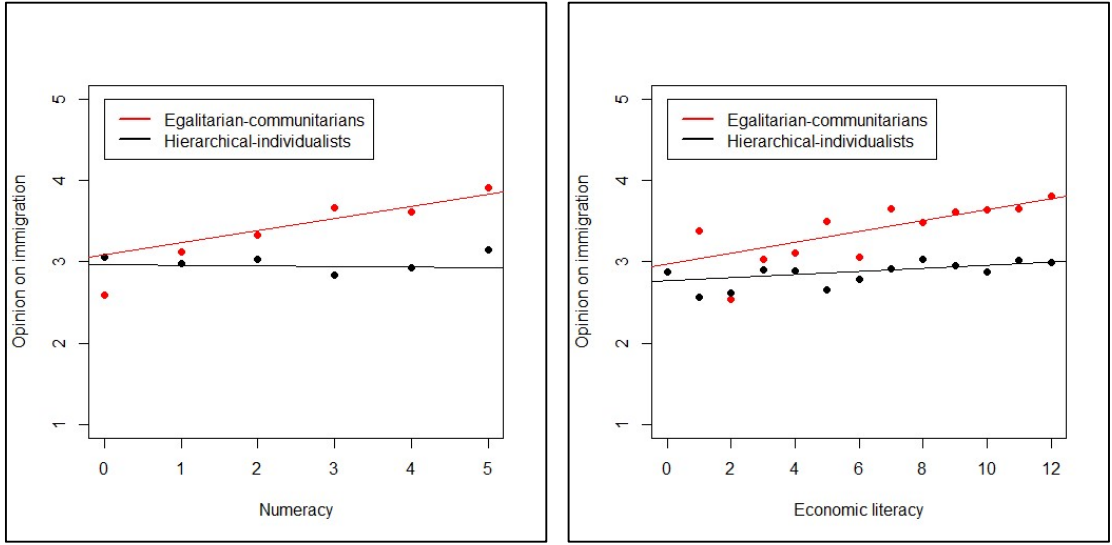

(a)  (b)

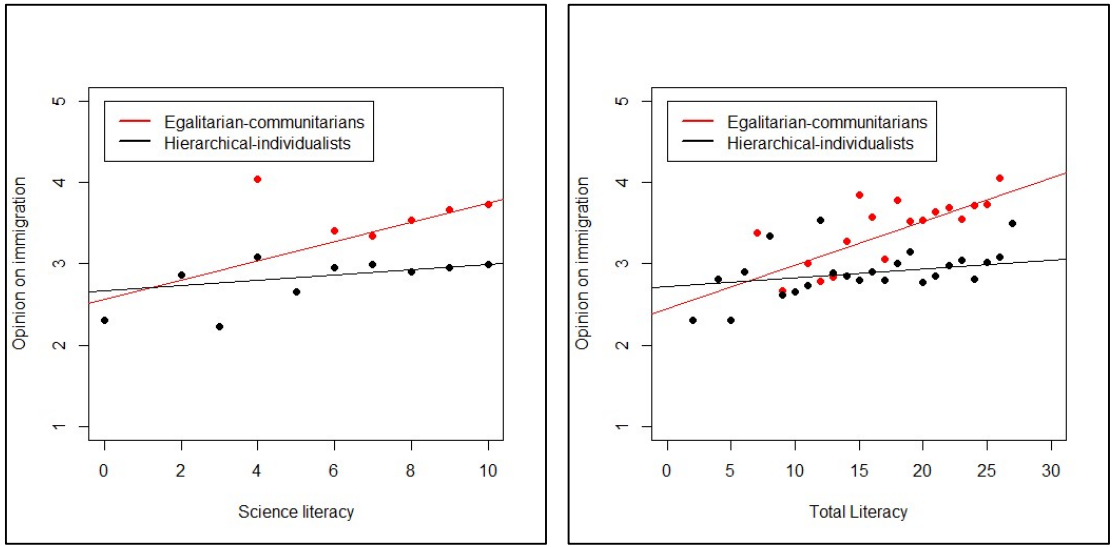

(c)  (d)

**FIGURE 1**. *Interaction among the explanatory x variable and the dummy z variable in the estimated models. The x variable is, respectively, numeracy (a), economic literacy (b), science literacy (c), and total literacy (d). Red lines represent the fitted values for egalitarian-communitarian individuals, while the black lines represent the fitted values for hierarchical-individualist individuals. Dots represent the mean values of opinion on immigration corresponding to the different levels assumed by the variable x for the egalitarian-communitarian individuals (red) and for the hierarchical-individualist individuals (black).*

# APPENDIX

## Opinion Toward Immigration Items

1. Immigrants improve Italian society by bringing new ideas and cultures.[1]
2. Italian culture is generally undermined by immigrants.[1]
3. Immigrants increase crime rates.[1]
4. Immigrants are generally good for Italian's economy.[1]
5. Immigrants take jobs away from people who were born in Italy.[1]
6. They should be entitled to have their children continue to qualify as Italian citizens if born in Italy.[1]
7. Legal immigrants to Italy who are not citizens should have the same rights as Italian citizens.[1]
8. The government spends too much money assisting immigrants.[1]
9. Refugees should be allowed to stay in Italy.[1]
10. Do you think the number of immigrants to America nowadays should be:[2]
11. In general, how risky do you consider the immigration phenomenon to be to Italian society as a whole?[3]
12. In general, how beneficial do you consider the immigration phenomenon to be to Italian society as a whole?[3]
13. Thinking about the immigration phenomenon in Italy makes you feel __________ emotions:[4]

[1] Extremely agree, agree, neither agree nor disagree, disagree, extremely disagree.
[2] Increased a lot, increased a little, remain the same as it is, reduced a little, reduced a lot.
[3] Extremely risky/beneficial, very risky/beneficial, reasonably risky/beneficial, not at all risky/beneficial.
[4] Very negative, a little negative, no emotion, a little positive, very positive.

## Numeracy Items

Q16 (CRT). If it takes five machines 5 minutes to make five widgets, how long would it take 100 machines to make 100 widgets? (100, 20, 10, or **5 minutes**)

Q15 (CRT). A notebook and a pen cost € 1.80 in total. The notebook costs € 1.00 more than the pen. How much does the pen cost? (€0.80, **€0.40**, €0.20, or € 1.00)[a]

Q3. In a lottery, the chance of winning a car is 1 in 1000. What percentage of tickets in that lottery wins a car? (**0.1%**, 1%, 10%, 0.5%)[b]

Q1. Imagine that we roll a fair, six-sided die 1000 times. Out of 1000 rolls, how many times do you think the die would come up as an even number? (300, 167, 150, or **500**)

Q9. If the chance of getting a disease is 20 out of 100, this would be the same as having _____ chance of getting the disease. (80%, 2%, **20%**, 40%)

*Note.* Correct answers are in boldface. The name of the items refers to the original test by Weller et al. (2013).
[a]The item was readapted to match the cultural context; the original item was about a bat and a ball.
[b]The item was readapted to match cultural context; in the original version, the lottery's name was ACME PUBLISHING SWEEPSTAKES.

# Science Literacy Items

1. The center of the Earth is very hot.

2. All radioactivity is man-made.

3. It is the father's gene that decides whether the baby is a boy or a girl.

4. Lasers work by focusing sound waves.

5. Electrons are smaller than atoms.

6. Antibiotics kill viruses as well as bacteria.

7. The continents on which we live have been moving their locations for millions of years and will continue to move in the future.

8. It is the Earth that goes around the Sun.

9. According to the theory of evolution, human beings, as we know them today, developed from earlier species of animals.

10. According to astronomers, the universe began with a big explosion.

# Supplementary File

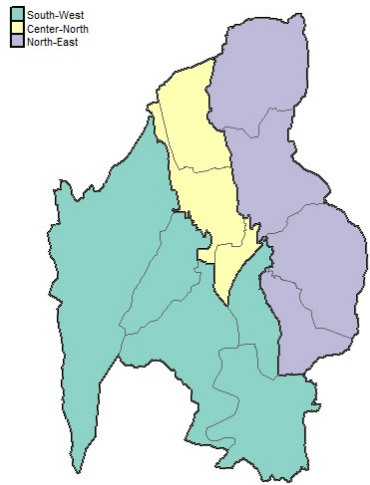

**FIGURE 1S**: *Map of the city of Trento (Italy). Black contours identify the 12 districts and the three macro-areas are identified by three colors. The yellow area corresponds to the old town of the city and it is the most populated zone.*

**Table 1S.** *Reference population (absolute values and percentages) stratified by gender (female/male), age groups (18-35, 36-55 and 56-80 years) and macro-area of residence (South-West, Center-North, North-East).*

|          |              | Absolute value | %    |
|----------|--------------|----------------|------|
| **Gender** | Female       | 46,261         | 51.4 |
|          | Male         | 43,790         | 48.6 |
|          | Total        | 90,051         | 100  |
| **Age**  | 18 – 35      | 23,678         | 26.3 |
|          | 36 – 55      | 33,555         | 44.8 |
|          | 56 – 80      | 32,818         | 31.1 |
|          | Total        | 90,051         | 100  |
| **Area** | South – West | 28,028         | 31.1 |
|          | Center – North | 40,359       | 44.8 |
|          | North – East | 21,664         | 24.1 |
|          | Total        | 90,051         | 100  |

**Table 2S.** *Selected sample (absolute values and percentages) stratified by gender (female/male), age groups (18-35, 36-55 and 56-80 years) and macro-area of residence (South-West, Center-North, North-East).*

|  |  | **Absolute value** | **%** |
|---|---|---|---|
| **Gender** | Female | 1,031 | 51.3 |
|  | Male | 977 | 48.7 |
|  | Total | 2,008 | 100 |
| **Age** | 18 – 35 | 530 | 26.4 |
|  | 36 – 55 | 747 | 37.2 |
|  | 56 – 80 | 731 | 36.4 |
|  | Total | 2,008 | 100 |
| **Area** | South – West | 625 | 31.1 |
|  | Center – North | 900 | 44.8 |
|  | North – East | 483 | 24.1 |
|  | Total | 2,008 | 100 |

**Table 3S.** *Response rate (absolute values and percentages) stratified by gender (female/male), age groups (18-35, 36-55 and 56-80 years) and macro-area of residence (South-West, Center-North, North-East).*

|  |  | Absolute value | % |
|---|---|---|---|
| **Gender** | Female | 270 | 26.2 |
|  | Male | 281 | 28.8 |
|  | Total | 551 | 100 |
| **Age** | 18 – 35 | 108 | 20.4 |
|  | 36 – 55 | 188 | 25.2 |
|  | 56 – 80 | 255 | 34.9 |
|  | Total | 551 | 100 |
| **Area** | South – West | 178 | 28.5 |
|  | Center – North | 234 | 26 |
|  | North – East | 139 | 28.8 |
|  | Total | 551 | 100 |